\documentclass[runningheads]{llncs}
\usepackage{subfig} 
\usepackage[table,xcdraw]{xcolor}
\usepackage{amsmath,amssymb}
\usepackage{hyperref}
\hypersetup{colorlinks=true}

\usepackage{graphicx}
\usepackage{amsmath}
\usepackage{amssymb}
\usepackage{booktabs}

\usepackage{amsfonts}

\usepackage{pifont}
\newcommand{\cmark}{\ding{51}}%
\newcommand{\xmark}{\ding{55}}%
\usepackage[figuresright]{rotating}
\usepackage{array}
\usepackage{booktabs}
\usepackage{makecell}
\usepackage{multirow}
\usepackage{color}
\usepackage{graphicx}
\usepackage{float}

\newcolumntype{x}[1]{>{\centering\arraybackslash}p{#1pt}}
\newcolumntype{a}[1]{>{\columncolor{verylightgray}\centering\arraybackslash}p{#1pt}}
\newcolumntype{y}[1]{>{\raggedright\arraybackslash}p{#1pt}}
\newcolumntype{z}[1]{>{\raggedleft\arraybackslash}p{#1pt}}

\newlength\savewidth\newcommand\shline{\noalign{\global\savewidth\arrayrulewidth\global\arrayrulewidth 1pt}\hline\noalign{\global\arrayrulewidth\savewidth}}
\newcommand{\tablestyle}[2]{\setlength{\tabcolsep}{#1}\renewcommand{\arraystretch}{#2}\centering\footnotesize}

\usepackage{sidecap}
\sidecaptionvpos{figure}{c}
\usepackage{algorithm}
\usepackage{algorithmic}

\definecolor{baselinecolor}{gray}{.9}

\newcommand{\etal}{\mbox{et al.}}
\newcommand{\eg}{\mbox{e.g.,\ }}
\newcommand{\ie}{\mbox{i.e.,\ }}

\begin{document}

\title{CateNorm: Categorical Normalization for Robust Medical Image Segmentation}
\titlerunning{Categorical Normalization}
%
\author{Junfei Xiao\inst{1} \and
Lequan Yu\inst{2} \and
Zongwei Zhou\inst{1} \and
Yutong Bai\inst{1} \and \\
Lei Xing\inst{3}  \and
Alan Yuille\inst{1} \and
Yuyin Zhou\inst{4} 
}
%


\authorrunning{J. Xiao et al.}
%

\institute{Johns Hopkins University \and
 The University of Hong Kong \and
 Stanford University \and
 UC Santa Cruz}

\maketitle              
\begin{abstract}

Batch normalization (BN) uniformly shifts and scales the activations based on the statistics of a batch of images. 
However, the intensity distribution of the background pixels often dominates the BN statistics because the background accounts for a large proportion of the entire image.
This paper focuses on enhancing BN with the intensity distribution of foreground pixels, the one that really matters for image segmentation.
We propose a new normalization strategy, named categorical normalization (CateNorm), to normalize the activations according to categorical statistics.
The categorical statistics are obtained by dynamically modulating specific regions in an image that belong to the foreground.
CateNorm demonstrates both precise and robust segmentation results across five public datasets obtained from different domains, covering complex and variable data distributions. 
It is attributable to the ability of CateNorm to capture domain-invariant information from multiple domains (institutions) of medical data.\\
Code is available at \href{https://github.com/lambert-x/CateNorm}{https://github.com/lambert-x/CateNorm}.

\end{abstract}
%
%
%
\section{Introduction}
\label{sec:introduction}

\begin{figure}
  \begin{minipage}[c]{0.49\textwidth}
    \includegraphics[width=\textwidth]{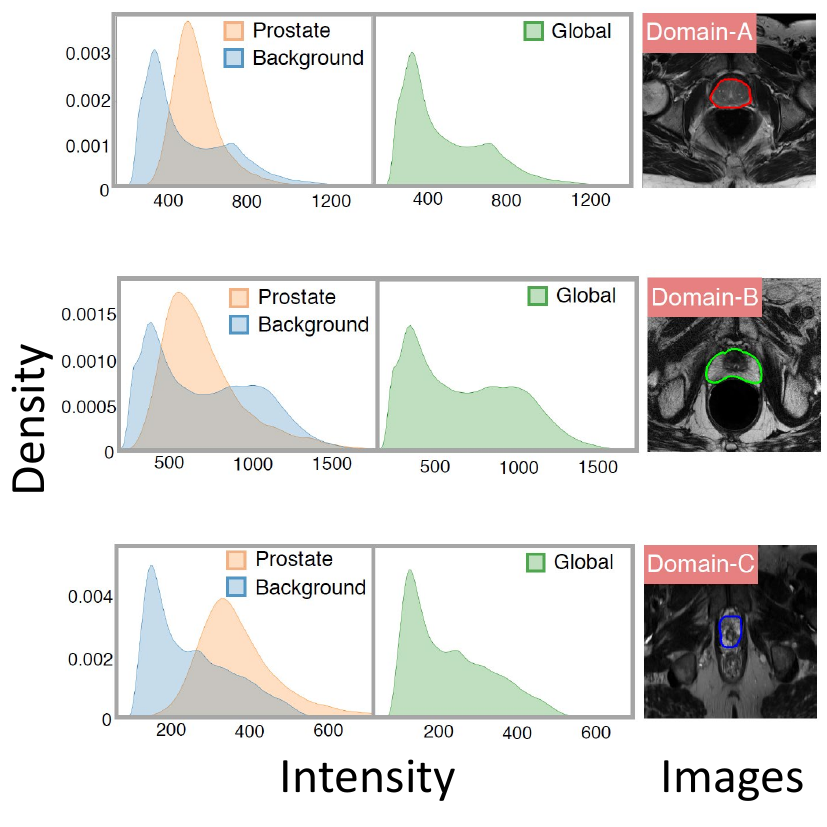}
  \end{minipage}
  \begin{minipage}[c]{0.49\textwidth}
  \caption{
   This paper addresses two limitations in conventional normalization strategies (\eg BN): 
   (\textit{i}) the intensity distribution of dominant classes, such as background, takes over the global statistics for normalization; 
   (\textit{ii}) the shift of intensity statistics across different domains is substantial (examples in Domains A--C).
   In response, we propose categorical normalization---instead of normalizing the intensity distribution based on the entire image, CateNorm normalizes the distribution per category.
   } \label{fig:classwise}
  \end{minipage}
\end{figure}


Normalization techniques are vital to accelerate and stabilize the network training procedure.
In addition to batch normalization (BN)~\cite{ioffe2015batch}, alternative techniques have been used, such as group normalization~\cite{wu2018group}, layer normalization~\cite{ba2016layer}, and instance normalization~\cite{ulyanov2017instance}. 
However, these normalization strategies are found to be suboptimal for medical image segmentation~\cite{liu2020ms} because they only estimate the statistics of the images as a whole, which can be easily biased towards the intensity distribution of dominant categories (see \figureautorefname~\ref{fig:classwise}).

To address this problem, this paper focuses on enhancing BN with the intensity distribution of foreground pixels---the one that matters for image segmentation.
We propose a new normalization strategy, named categorical normalization, named \textbf{CateNorm}, to normalize the activations according to categorical statistics.
We introduce CateNorm to the U-Net architecture by enforcing the normalization layers to modulate foreground regions (\eg pancreas, stomach) differently.
Specifically, two parallel and complementary schemes are adopted for normalizing the activations---one is the conventional BN to capture the statistics of a batch of images; the other is the proposed CateNorm to integrate the statistics of specific regions with the guidance from the learned categories.
As CateNorm conditions on the categorical masks (not available in the inference phase), we hereby introduce a simple yet effective two-stage training strategy: 
concretely, we first train exclusively using BN for generating the categorical masks and then feed them to CateNorm for updating the semantic-related modulating parameters.

Our extensive experiments on five datasets show that the integration of categorical statistics into normalization strategies can better capture the domain-invariant information from different domain data, thereby robustifying the learned medical representation.
Compared with the existing normalization techniques \cite{ba2016layer,ioffe2015batch,ulyanov2017instance,wu2018group}, CateNorm achieves more precise and robust segmentation results in various applications, including multi-organ segmentation from CT images and prostate segmentation from MRI images. 
CateNorm also consistently improves over the previous state of the arts for multi-domain data.
This result suggests that CateNorm not only extracts more discriminative features but also compensates for the statistical bias in small distribution shifts and domain gaps.

Our contributions are three-fold:
(1) a novel normalization to integrate categorical statistics with BN's general statistics, surpassing existing normalization techniques;
(2) a light CateNorm residual block for the U-Net encoder, yielding a prominent performance gain with negligible additional parameters;
(3) CateNorm achieves better robustness in complex and variable data distributions.

\section{Related Work}
\label{sec:related_work}

Normalization is one of the keys to the success of deep networks. As the most commonly used normalization technique, batch normalization (BN)~\cite{ioffe2015batch} enables training with larger learning rates and greatly mitigates general gradient issues. 
Besides, alternative normalization strategies have also been designed for specific scenarios: layer normalization (LN)~\cite{ba2016layer} for recurrent neural networks, instance normalization (IN)~\cite{ulyanov2017instance} for style transfer, group normalization (GN)~\cite{wu2018group} for small-batch training, etc.
Beyond the natural image domain, these normalization strategies have also been successfully applied to medical applications.
For example, Kao~\etal~\cite{kao2018brain} apply GN for brain tumor segmentation; Isensee~\etal~\cite{isensee2018no} apply IN in a self-adaptive framework for various segmentation tasks; Chen~\etal~\cite{chen2021transunet} apply LN in Transformers for abdominal multi-organ segmentation. 
A detailed comparison among different normalization strategies for medical semantic segmentation and cross-modality synthesis has been summarized in Zhou~\etal~\cite{zhou2019normalization} and Hu~\etal~\cite{hu2019cross}.
To offer stronger affine transformation, the latest strategies~\cite{dumoulin2016learned,huang2017arbitrary} utilize external data to denormalize the features. As class information could be ``washed away'' with previous strategies,  SPADE~\cite{park2019semantic} directly uses class masks to guide the normalization.  
In addition, for domain adaptation and multi-domain learning, other methods propose to modify BN by modulating~\cite{li2016revisiting} or calculating domain-specific~\cite{chang2019domain,liu2020ms} statistics.
Unlike existing strategies, this paper proposes a novel categorical normalization strategy for incorporating different types of statistics.


\section{Categorical Normalization (CateNorm)}
\label{sec:method}

\smallskip\noindent\textbf{Proof of concept:} 
A simple experiment is designed to illustrate how categorical statistics can better mitigate distribution shifts across domains.
We align the input image distributions based on the categorical statistics obtained from different datasets.
Then we jointly train these heterogeneous datasets on the aligned input images. With class-wise distribution alignment, the model gains an improvement of 0.7\% (91.4\% vs. 90.7\%).
This suggests that aligning the data distribution based on class-wise statistics can better mitigate domain shifts between datasets and therefore achieve better results in the joint training setting.
This interesting observation further motivates us to design normalization strategies to further leverage local statistics during the learning process.

We devise a two-stage training paradigm. In the first stage, the network is trained with BN to generate the class masks, which are later fed to the second stage to normalize the distribution based on each category. \figureautorefname~\ref{fig:overview} presents the overall pipeline of our approach.
In the following, we will first introduce our CateNorm Residual Block (CNRB), following with the overall training and testing pipeline of CateNorm.

\smallskip\noindent\textbf{CateNorm Residual Block:}
In addition to BN in residual blocks~\cite{he2016deep}, we introduce categorical normalization scheme while keeping all other components the same (see \figureautorefname~\ref{fig:overview}(b)).
Let $x$ with the spatial resolution of $H \times W$ denotes the input feature map, with a batch of $N$ samples and $C$ number of channels. 
For the $n$-th sample at the $c$-th channel, $x_{n, c, i, j}$ denotes the associated activation value at spatial location $(i, j)$. 
Then a BN branch and a CateNorm branch are used for fully leveraging the general and categorical statistics, respectively.
BN~\cite{ioffe2015batch} estimates the statistics of a batch of images and then applies affine transformation with learnable parameters $\gamma^{\textup{BN}}_{c}$ and $\beta^{\textup{BN}}_{c}$ at the $c$-th channel. 
Specifically, we first compute the mean and standard deviation $\mu_{c}$ and $\sigma_{c}$ as follows: $\mu_{c} =\frac{1}{N H W} \sum_{n, i, j} x_{n, c, i, j}$, 
$\sigma_{c} =\sqrt{\frac{1}{N H W} \sum_{n, i, j}\left(\left(x_{n, c, i, j}\right)^{2}-\left(\mu_{c}\right)^{2} \right ) + \epsilon}$,
where $\epsilon$ denotes a small constant for avoiding invalid 
denominators.
And the associated activation map can be then computed as:
\begin{equation}
\centering
\gamma^{\textup{BN}}_{c} \cdot \frac{x_{n, c}-\mu_{c}}{\sigma_{c} }+\beta^{\textup{BN}}_{c}.
\label{eq:BN}
\end{equation}

\begin{figure}[t]
\begin{center}
\includegraphics[width=1.0\textwidth]{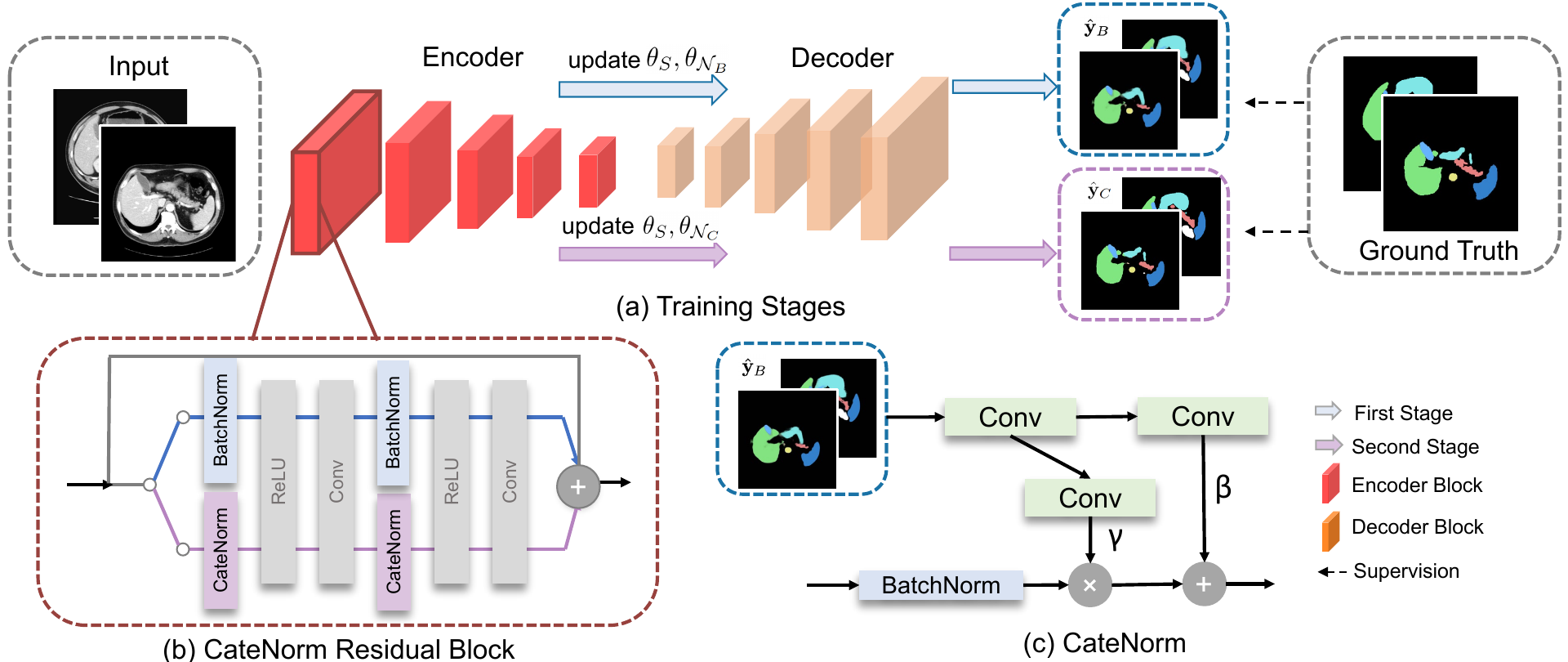}
\end{center}
  \caption{Framework overview: (a) architecture \& training stages of U-Net using BN and CateNorm; (b) design of CateNorm residual block; (c) design of CateNorm.}
\label{fig:overview}
\end{figure}

CateNorm serves as an addition normalization to integrate categorical statistics.
The key difference, compared with BN, is that CateNorm provides a spatially-variant affine transformation which is learned from the corresponding class mask for modulating the activations.
Therefore, the modulation parameters $\gamma^{\textup{CateNorm}}(\mathbf{m})$ and $\beta^{\textup{CateNorm}}(\mathbf{m})$ are no longer $C$-dimensional vectors as aforementioned, but tensors with spatial resolution $H \times W$. 
Here $\gamma^{\textup{CateNorm}}(\cdot)$ and $\beta^{\textup{CateNorm}}(\cdot)$ are learnable functions that act on the given class mask $\mathbf{m}$.

The activations of the CateNorm branch are computed as:
\begin{equation}
\centering
\gamma^{\textup{CateNorm}}_{c, i, j}(\mathbf{m}) \frac{x_{n, c, i, j}-\mu_{c}}{\sigma_{c}}+\beta^{\textup{CateNorm}}_{c, i, j}(\mathbf{m}).
\label{eq:CateNorm}
\end{equation}

Inspired by SPADE~\cite{park2019semantic}, we use a lightweight two-layer convolutional network to learn the modulation functions $\gamma^{\textup{CateNorm}}(\cdot)$ and $\beta^{\textup{CateNorm}}(\cdot)$, where the first layer is set to output features of $C / 2$ number of channels.

\smallskip\noindent\textbf{Training CateNorm:} We inject CateNorm into popular segmentation models, such as U-Net~\cite{ronneberger2015u} and DeepLabV3+~\cite{chen2018encoder}. The model can be denoted by $\mathcal{F}(\cdot;\theta)$ parameterized by $\theta = \{\theta_s, \theta_{\mathcal{N}_{B}}, \theta_{\mathcal{N}_{C}}\}$.  $\theta_{\mathcal{N}_{B}}$ denote the learnable modulating parameters used in the BN branch and $\theta_{\mathcal{N}_{C}}$ is the learning modulation function that requires the corresponding class mask as an input; the subscripts $\mathcal{N}_{B}$ and $\mathcal{N}_{C}$ here denote BN and CateNorm. $\theta_s$ stands for all other network parameters.

In the first stage, we exclusive train the model through the BN branch, \emph{i.e.}, only $\theta_s$ and $\theta_{\mathcal{N}_{B}}$ are updated.
The goal of the first stage is not only to leverage the global statistics for accelerating training but  also to provide class information for learning CateNorm parameters in the second stage. 
Specifically, the class mask generated from the first stage $\hat{\mathbf{y}}_B$ can be written as:
\begin{align}
\hat{\mathbf{y}}_B &= \mathcal{F}(\theta_{S},\theta_{\mathcal{N}_{B}}; \mathbf{x}).
\label{eq:bn_output}
\end{align}

In the second training stage, we train exclusively on the CateNorm branch for exploiting the local statistics, \emph{i.e.}, only update network parameters $\theta_s$ and the learnable function $\theta_{\mathcal{N}_{C}}(\cdot)$.
Given the class mask $\hat{\mathbf{y}}_B$, the normalization parameters $\theta_{\mathcal{N}_{C}}(\hat{\mathbf{y}}_B)$ can be then computed via Equation~\eqref{eq:CateNorm}, and the predicted mask in the second stage $\hat{\mathbf{y}}_C$ can be written as:
\begin{equation}
\hat{\mathbf{y}}_C = \mathcal{F}(\theta_{S},\theta_{\mathcal{N}_{C}}(\hat{\mathbf{y}}_B); \mathbf{x}),
\label{eq:catenorm_output}
\end{equation}
where $\hat{\mathbf{y}}_C$ denotes the softmax probability map from the CateNorm branch. With auxiliary class information integrated, this training step aims at enhancing discriminative features, which leads to more accurate and robust segmentation 

\smallskip\noindent\textbf{Overall training objective:} Given a pair of probability prediction $\hat{\mathbf{y}}$ and the associated ground truth $\mathbf{y}
\in \mathbb{L}^{D}$, the Dice loss and the cross entropy loss are:
$\mathcal{L}_{Dice}(\mathbf{y},\hat{\mathbf{y}}) = \frac{1}{|\mathbb{L}|}\sum_l [1-\frac{2 \sum_{i,j,l} y_{i,j,l}\cdot \hat{y}_{i,j,l}}{\sum_{i,j,l} (y_{i,j,l}^2 + \hat{y}_{i,j,l}^2)}]$,
$\mathcal{L}_{CE}(\mathbf{y},\hat{\mathbf{y}}) = -\frac{1}{D \cdot |\mathbb{L}|}\sum_{i,j,l} y_{i,j,l}\cdot \log (\hat{y}_{i,j,l})$,
where $\hat{y}_{i,j,l}$ is the output probability of the $l$-th class ($l \in \mathbb{L}$) of  at spatial location $i,j$. In our loss function, we use a weighted sum of these two losses, which can be written as: $\mathcal{L}(\mathbf{y},\hat{\mathbf{y}}) = \lambda \mathcal{L}_{Dice}(\mathbf{y},\hat{\mathbf{y}}) + (1 - \lambda) \mathcal{L}_{CE}(\mathbf{y},\hat{\mathbf{y}})$,
where $\lambda$ is the balance parameter. Therefore, our overall training objective over these two stages is:
\begin{align}
&\mathcal{L}_{total} = \alpha \mathcal{L}(\mathbf{y},\hat{\mathbf{y}}_B) + (1 - \alpha) \mathcal{L}(\mathbf{y},\hat{\mathbf{y}}_C)
,
\end{align}
where $\alpha$ is set as 1 in the first stage and 0 in the second stage, for updating $\{\theta_s,\theta_{\mathcal{N}_{B}}\}$ and $\{\theta_s, \theta_{\mathcal{N}_{C}}\}$ alternately.
In the testing phase, the final prediction is obtained by forwarding twice with Equations~\eqref{eq:bn_output} \&~\eqref{eq:catenorm_output} sequentially.
The whole training procedure is summarized in Appendix Algorithm \ref{alg:training}.

\section{Experiments}

\subsection{Dataset and Benchmark}

\smallskip\noindent\textbf{Prostate segmentation datasets:} Following Liu~\etal~\cite{liu2020ms}, we use prostate T2-weighted MRI collected from three different domains. (1) 30 samples from  Radboud University Nijmegen
Medical Centre;
(2) 30 samples from Boston Medical Center.
Both (1) and (2) are available from NCI-ISBI 2013
challenge (ISBI 13) dataset~\cite{ISBI2013}, therefore are denoted as ``ISBN-R'' and ``ISBN-B''.
(3) 19 samples from Initiative for Collaborative Computer Vision Benchmarking
(I2CVB) dataset~\cite{lemaitre2015computer}, denoted as ``I2CVB''.
Details of their acquisition protocols are included in Appendix \S\ref{sec:appendix_protocols}.

\smallskip\noindent\textbf{Abdominal multi-organ segmentation datasets:} We use abdominal CT images of two different domains. 
(1) 30 training cases from the Beyond the Cranial Vault (BTCV) dataset~\cite{landman20152015};
(2) 41 cases from the Cancer Image Archive (TCIA) Pancreas-CT dataset \cite{clark2013cancer,TCIA_Data,roth2015deeporgan}, where the multi-class annotation can be acquired from Gibson~\etal~\cite{gibson_eli_2018_1169361}.
For single-domain experimental settings, 8 organs (spleen, left kidney, right kidney, gallbladder, pancreas, liver, stomach, and aorta) are evaluated following the settings in Fu~\etal~\cite{fu2020domain} and Chen~\etal~\cite{chen2021transunet}.
For multi-domain experiments, right kidney and aorta are excluded and we evaluate the remaining 6 organs which are labeled in both datasets.

\begin{table*}[t]
\centering
    \caption{
        CateNorm vs. three other normalization strategies under the single-domain setting. The performance is measured by Dice score (\%). CN denotes our categorical normalization. We report the average performance on five-fold cross-validation, along with the statistic analysis (*$p<$0.5, **$p<$0.1, ***$p<$0.05) between the best and second best methods.
    }
    \label{tab:compare_prostate}
    \begin{tabular}{p{0.22\linewidth}p{0.13\linewidth}|p{0.15\linewidth}|p{0.15\linewidth}p{0.15\linewidth}p{0.15\linewidth}}
        Method  & Norm & BTCV  & ISBN-R & ISBN-B & I2CVB \\ 
        \shline
        Baseline  & BN   & 75.56         & 88.69 & 85.20  & 87.21 \\
        Baseline  & BN   & 77.83         & 89.13 & 85.99 & 88.22 \\ \hline 
        Baseline  & BN   & 78.98         & 90.08 & 87.22 & 88.99 \\
        Baseline  & IN   & 78.96         & 89.05 & 88.37 & 88.98  \\
        Baseline  & GN   & 78.50    & 89.15 & \textbf{88.61} & 89.16 \\ \hline
        Ours (block 1)   & CN & 79.33 &  91.36          & {88.55} & 89.31   \\
        Ours (block 1-4)  & CN & \textbf{80.37}** & \textbf{91.41}** & 87.57 & \textbf{89.79}** \\
    \end{tabular}
    
\end{table*}

\begin{table*}[t]
\centering
    \caption{
        CateNorm vs. three other methods under the multi-domain setting. CN denotes our categorical normalization. We report the average performance on five-fold cross-validation, along with the statistic analysis (*$p<$0.5, **$p<$0.1, ***$p<$0.05) between ours (bolded) and the best previous method (DSBN~\cite{chang2019domain}).
    }
    \label{tab:multisite_prostate}
    \begin{tabular}{p{0.2\linewidth}p{0.09\linewidth}|p{0.13\linewidth}p{0.13\linewidth}|p{0.13\linewidth}p{0.13\linewidth}p{0.13\linewidth}}
        Method         & Norm     & BTCV   & TCIA    & ISBN-R         & ISBN-B         & I2CVB                \\ 
        \shline
        Baseline       & BN       & 82.64 & 87.33 & 91.50          & 90.46          & 90.34                \\
        DSBN          & BN       & 82.67& 87.83 & 91.98          & 90.22          & 90.10                \\
        MS-Net        & BN       & 82.17 & 87.85 & 91.93          & 90.30          & 89.89              \\ 
        \hline
        Ours (block1)   & CN & 83.28 & 88.22 & 92.12          & 90.76          & 90.33               \\
        Ours (block1-4) & CN & \textbf{83.45}* & \textbf{88.38}** & \textbf{92.47}*** & \textbf{91.17}*** & \textbf{90.79}* \\
    \end{tabular}
    
\end{table*}

\subsection{Robust Performance on Single- \& Multi-Domain Data}
\label{sec:comparison_sota}
We compare the proposed CateNorm strategy with various normalization
methods, including BN~\cite{ioffe2015batch}, IN~\cite{ulyanov2017instance}, and GN~\cite{wu2018group}. 
In addition, for multi-domain settings, we also compare with state-of-the-art multi-domain learning approaches, including DSBN~\cite{chang2019domain}, MS-Net~\cite{liu2020ms}.
To ensure a fair comparison, we implement UNet with residual blocks in the encoder \cite{liu2020ms,yu2017volumetric} for all baseline methods.

\smallskip\noindent\textbf{Single-domain results:} As shown in \tableautorefname~\ref{tab:compare_prostate}, even with strong data augmentation, our CateNorm still yield a solid performance gain on all four datasets.
For instance, on the prostate dataset ``ISBN-R'' and the multi-organ segmentation dataset, CateNorm outperforms BN by a large margin of $1.33\%$ and $1.39\%$ in average Dice. 
While different normalization methods (\eg GN, IN) may behave similarly, 
our CateNorm consistently achieves better results compared with all other methods.
We also compare two different configurations of CateNorm: 1) \emph{CateNorm (block 1)} only replaces the first encoder block with CNRB;
2) \emph{CateNorm (block 1-4)} replaces all of the first four encoder blocks with CNRBs.
For prostate segmentation, we find that
both variants show a solid improvement while CateNorm \emph{(block 1)} with only few additional parameters performs similarly as CateNorm \emph{(block 1-4)} (\ie $89.74\%$ vs. $89.59\%$).
On the contrary, for multi-organ segmentation, \emph{block 1} demonstrates inferior results than \emph{block 1-4} (\ie $79.33\%$ vs. $80.37\%$).
This suggests that CNRB can bring additional benefits for complex tasks such as multi-organ segmentation.
For the relatively simpler binary segmentation task, \emph{block 1} might be enough to learn a good model, therefore using more CNRBs does not lead to further performance gain.
A detailed study regarding where to add CNRBs has been illustrated in \S\ref{analytical study}.

\smallskip\noindent\textbf{Multi-domain segmentation results:} We evaluate our method using the same multi-domain setting as in Liu~\etal~\cite{liu2020ms}. 
To demonstrate the effectiveness of CateNorm, we compare the performance with the baseline and state-of-the-art multi-domain learning methods (\ie DSBN, MS-Net) on three prostate segmentation datasets and two multi-organ segmentation datasets. 
As shown in \tableautorefname~\ref{tab:multisite_prostate}, 
under strong data augmentation (\eg rotation, flipping), DSBN and MS-Net do not yield improvements anymore, while
our method still secures a reasonable improvement compared to the baseline.
For instance, \emph{block 1-4} outperforms the baseline by $0.81\%$ and $1.05\%$ in average Dice on the BTCV and TCIA dataset, respectively.
This indicates that, unlike previous methods which customize the normalization layers for different domains~\cite{chang2019domain,liu2020ms}, our CateNorm can better extract domain-invariant information in the face of a more complex and variable data distribution.
Meanwhile, it is also worth mentioning that our approach is complementary to previously domain-specific normalization methods. 

Unlike the single-domain setting, \emph{block 1-4} outperforms \emph{block 1} for both prostate segmentation and multi-organ segmentation.
We conjecture that this is due to that given a more complex data distribution, more CNRBs can bring additional benefits by imposing semantic guidance on the encoder more densely. 
More importantly, our CateNorm is flexible to many popular segmentation architectures such as DeepLabV3+ (see Appendix \tableautorefname~\ref{tab:deeplab}).
Moreover, CateNorm shows great robustness in partially annotated scenarios---detailed studies are provided in Appendix \figureautorefname~\ref{fig:less_classes}. Qualitative results are in Appendix \figureautorefname~\ref{fig:Qualitative}.

\subsection{Discussion and Ablation Study}
\label{analytical study}


\begin{table}[t]
\caption{Parameter sharing for the two-stage normalization (Dice Score in \%)}
\subfloat[Sharing vs. no-sharing
\label{tab:network_share}
]{%
\begin{minipage}{0.45\linewidth}
\begin{center}
\footnotesize
\tablestyle{1pt}{1.0}
\begin{tabular}{@{}l|cc|c}
Method       & Sharing & \#Params       & Dice                      \\ 
\shline
U-Net     & --      & 1.0$\times$   & 90.08                   \\
W-Net & \xmark  & 2.073$\times$ & 90.36                   \\
\hline
\textbf{Ours(blk1)}        & \cmark  & 1.001$\times$ & 91.36  \\ 
\textbf{Ours(blk1-4)}      & \cmark  & 1.073$\times$ & \textbf{91.41}  \\
\end{tabular}
\end{center}
\end{minipage}
}
\hspace{1em}
\subfloat[Mutual benefits of BN and CateNorm.
\label{tab:both_improved}
]{%
\begin{minipage}{0.5\linewidth}
\begin{center}
\footnotesize
\tablestyle{1pt}{1.0}
\begin{tabular}{l|c|cc}
                      & Forward  & Prostate       & Abdominal      \\ 
                      \shline
Baseline              & BN    & 90.76          & 84.98          \\
\hline
\multirow{2}{*}{\textbf{Ours}} & BN    & 91.00          & 85.39          \\
                      & CN & \textbf{91.48} & \textbf{85.92}
                      \\
\end{tabular}
\end{center}
\end{minipage}
}
     \label{tab:sharing_two_tables}
\end{table}

\smallskip\noindent\textbf{Parameter sharing for the two-stage normalization:} To prove the necessity of network sharing except the normalization layers, we also implement our method by using different network parameters $\theta_s$ in the two training stages, similar to W-Net~\cite{xia2017w}.
In this implementation, CateNorm and BN are deployed in two independent sub-networks, which are simply concatenated for training and testing.
As shown in \tableautorefname~\ref{tab:network_share}, our method performs much better than W-Net with only about 50\% of the parameters.
Besides, even only comparing the results in the first stage where only BN is used during inference, as shown in \tableautorefname~\ref{tab:both_improved}, our approach still outperforms the baseline by $0.24\%$ and $0.41\%$ in average Dice.
Then in the second stage where CateNorm is used for inference, the performance can be further improved by $0.48\%$ and $0.53\%$.
This indicates that by sharing the rest of the network parameters, the two normalization schemes can mutually benefit each other by leveraging both general and categorical statistics.

\smallskip\noindent\textbf{Adding CateNorm to the earlier encoder:} 
As shown in \S\ref{sec:comparison_sota}, adding more CateNorm in the U-Net encoder can benefit both prostate and multi-organ segmentation, especially under the multi-domain setting.
This observation motivates us to further investigate where to add CateNorm, as this can help us design better configurations of CateNorm which achieve higher performance without incurring much computation cost.
In our experiments, U-Net consists of 5 encoder blocks and 5 decoder blocks. By varying the position to add the CateNorm from blocks 1 to 10, we compare the average Dice score on the BTCV dataset.
As shown in Appendix \figureautorefname~\ref{fig:where_abdominal}, CateNorm blocks are preferred to be set in early blocks (encoder). 

\smallskip\noindent\textbf{Visualizing activations in CateNorm:}
Appendix \figureautorefname~\ref{fig:Visualize_norm} visualizes the learned $\gamma^{\textup{CateNorm}}$ and $\beta^{\textup{CateNorm}}$ on different channels of the intermediate CateNorm layers during the second forward. 
With prior class information as guidance, CateNorm can modulate spatially-adaptive parameters.
Such spatial-wise modulation can be complementary to the channel-wise modulation accomplished by BN, and derives more discriminative features that benefit segmentation.


\section{Conclusion}
\label{sec:conclusion}

We have presented a new normalization strategy, named CateNorm, which complementarily enhance the categorical statistics in BN for robust medical image segmentation. Our CateNorm can be used as an add-on to existing segmentation architectures, such as U-Net and DeepLabV3+. Compared with existing normalization strategies, CateNorm consistently achieves superior results, even with complex and variable data distributions. We believe that the proposed normalization strategy could also improve natural image segmentation and plan to explore it in the future work. 

\medskip
\noindent\textbf{Acknowledgments:} This work was supported by the Lustgarten Foundation for Pancreatic Cancer Research. We also thank Quande Liu for the discussion.

\newpage
{\small
\bibliographystyle{splncs04}
\bibliography{ref}}

\clearpage
\appendix

\section{Details of Aligning Input Distribution Algorithm}
\label{sec:align}

Assume that we have $N$ source domains $S_1, S_2, S_3, ..., S_N$, with $M_1, M_2, M_3, ..., M_N$ examples respectively, where the $i$-th domain source domain $S_i$ consists of an image set $\{\mathbf{x}_{i,j}\in \mathbb{R}^{D_{i,j}}\}_{j=1,...,M_i}$ as well as their associated annotations.
Our goal is to align the image distributions of these source domains with the target domain $T$ based on the class-wise (region-wise) statistics. 
The algorithm can be illustrated as the following steps:
\vspace{2mm}

\noindent
\textbf{Step 1: Calculate class-wise statistics of each case}

Firstly, we calculate the mean and standard deviation of each case in both the source domain and the target domain.
\begin{align}
\centering
\mu_{i,j}^{c} &= \frac{\sum_{k=1}^{|D^c_{i,j}|}\mathbf{x}^c_{i,j,k}}{|D^c_{i,j}|} , \\
\sigma_{i,j}^{c} &= \sqrt{\frac{1}{ |D^{c}_{i,j}|} \sum_{k=1}^{|D^{c}_{i,j}|}(\mathbf{x}^c_{i,j,k} - \mu_{i,j}^{c})^{2}},
\end{align}
where $\mathbf{x}_{i,j}^{c}$ denotes the pixels which belong to the $c$-th class (region) in image $\mathbf{x}_{i,j}$, with the number of pixels denoted as $|D^c_{i,j}|$. 
As a special case, $i=T$ indicates the target domain.
\vspace{2mm}

\noindent
\textbf{Step 2: Estimate aligned (new) class-wise statistics  }

Next, we calculate the mean of the statistics over all examples obtained in each domain as follows:
\begin{align}
\centering
\label{eq:stat_of_mean}
\Bar{\mu}_{i}^{c} &=\frac{\sum_{j=1}^{M_i}\mu_{i,j}^{c}}{M_i}, \\
\Bar{\sigma}_{i}^{c} &=\frac{\sum_{j=1}^{M_i}\sigma_{i,j}^{c}}{M_i}.
\end{align}

Based on the $\Bar{\mu}_{i}^{c}$, we now estimate the new class-wise mean $\Tilde{\mu}_{i,j}$ for each case of the source domain $S_i$ as follows:

\begin{align}
\begin{split}
\Tilde{\mu}_{i,j}^{c} &=\frac{\mu_{i,j}^{c} - \Bar{\mu}_{i}^{c}}{\sqrt{\frac{\sum_{j=1}^{M_i}(\mu_{i,j}^{c} - \Bar{\mu}_{i}^{c})^2}{M_i}}} \cdot  \sqrt{\frac{\sum_{j=1}^{M_T}(\mu_{T,j}^{c} - \Bar{\mu}_{T}^{c})^2}{M_T}} + \Bar{\mu}_{T}^{c}, \\
\end{split}
\end{align}
where $M_T$ denotes the number of cases in the target domain $T$.
Similarly, the new standard deviation $\Tilde{\sigma}_{i,j}$ can be computed by:

\begin{align}
\begin{split}
\Tilde{\sigma}_{i,j}^{c} &=\frac{\sigma_{i,j}^{c} - \Bar{\sigma}_{i}^{c}}{\sqrt{\frac{\sum_{j=1}^{M_i}(\sigma_{i,j}^{c} - \Bar{\sigma}_{i}^{c})^2}{M_i}}} \cdot  \sqrt{\frac{\sum_{j=1}^{M_T}(\sigma_{T,j}^{c} - \Bar{\sigma}_{T}^{c})^2}{M_T}} + \Bar{\sigma}_{T}^{c}. \\
\end{split}
\end{align}

\noindent
\textbf{Step 3: Align each case with the estimated statistics}

Based on the computed new mean and standard deviation $\Tilde{\mu}_{i,j}$, $\Tilde{\sigma}_{i,j}$, the aligned image $\Tilde{\mathbf{x}}_{i,j}$ can be computed as:
\begin{align}
\centering
\Tilde{\mathbf{x}}_{i,j}^{c} &=\frac{\mathbf{x}_{i,j}^{c} - \mu_{i,j}^{c}}{\sigma_{i,j}^{c}} \cdot \Tilde{\sigma}_{i,j}^{c} + \Tilde{\mu}_{i,j}^{c}.
\end{align}

\section{Implementation Details}

\begin{table}[htbp]
\tablestyle{6pt}{1.1}
\caption{\textbf{Data Preprocessing.}}
\begin{tabular}{c|ll}
Step & Prostate & Abdominal \\
\shline
1 & Center-cropping &  Window range Clipping [-125, 275] \\
2 & Out-of-mask slice cropping & Out-of-mask slice cropping \\
3 & Resizing & Resizing \\
4 & Z-score Normalization & Z-score Normalization

\end{tabular}

\label{tab:data_preprocess} \vspace{-.5em}
\end{table}

\begin{table}[htbp]
\caption{\textbf{Experimental Setting.}}
\tablestyle{6pt}{1.1}
\begin{tabular}{l|y{140}}
config & value \\
\shline
training iterations & 9000 \\
optimizer & Adam \\
initial learning rate & 1e-3 \\
optimizer momentum & $\beta_1, \beta_2{=}0.9, 0.999$ \\
batch size & 4 (single) 6 (multi) \\
{learning rate schedule}  & plateau scheduler  \\
Dice/CE balance factor $\lambda$ & 0.5 (abdominal) 1.0 (prostate) \\

\multirow{2}{*}{augmentation}  & horizontal flipping (prostate only) + random rotation \\
validation strategy & 5-fold \\
evaluation metric & Dice Score (\%) and ASD (mm)
\end{tabular}

\label{tab:training_setting} \vspace{-.5em}
\end{table}

\clearpage

\section{Details of the Prostate Datasets}
\label{sec:appendix_protocols}
\begin{table}[htbp]
\footnotesize
\begin{center}
\caption{Details of the 3 prostate segmentation datasets.}
\begin{tabular}{c|ccccc}
Dataset   & \#Cases &
\makecell{Field {strength} (T)} & \makecell{Resolution\\(in/through plane)}   &
Manufacturer\\  \shline
ISBN-R  & 30    & 3     & 0.6-0.625/3.6-4  & Siemens \\ 
ISBN-B   & 30    & 1.5   & 0.4/3  & Philips \\
I2CVB  & 19    & 3     & 0.67-0.79/1.25    & Siemens \\ 
\end{tabular}

\end{center}
\label{tab:prostate_details}
\end{table}


\section{Training procedure of CateNorm}
\begin{algorithm}[ht]
\caption{Training procedure of CateNorm}
\label{alg:training}
\begin{algorithmic}[1]
\REQUIRE Images and labels $\mathbf{x}$, $\mathbf{y}$; \\
~~~~~~~Network parameters $\theta = \{\theta_{S}, \theta_{\mathcal{N}_{B}}, \theta_{\mathcal{N}_{C}}\}$; \\
~~~~~~~Training iterations $\tau$;
\ENSURE Optimized parameters $\theta_{S}$, $\theta_{\mathcal{N}_{B}}$, $\theta_{\mathcal{N}_{C}}$;

    \STATE t $\leftarrow$ 0;
    \STATE Initialize $\theta_{S}$, $\theta_{\mathcal{N}_{B}}$ with the pretrained model and randomly initialize $\theta_{\mathcal{N}_{C}}$;
    \WHILE{$t<\tau$}
        \STATE Compute the class mask $\hat{\mathbf{y}}_B$;
        \STATE $\alpha \leftarrow 1$;
        \STATE Update ${\theta_s, \theta_{\mathcal{N}_{B}}}\leftarrow \mathop{\mathrm{min}_{\theta_s, \theta_{\mathcal{N}_{B}}}}
        {\mathcal{L}_{total}}$;
        \STATE Detach $\hat{\mathbf{y}}_B$ from gradient calculation;
        \STATE Compute the class mask $\hat{\mathbf{y}}_C$;
        \STATE $\alpha \leftarrow 0$;
        \STATE Update ${\theta_s, \theta_{\mathcal{N}_{C}}(\cdot)}\leftarrow \mathop{\mathrm{min}_{\theta_s, \theta_{\mathcal{N}_{C}}}}
        {\mathcal{L}_{total}}$;
        
        \STATE $t \leftarrow t + 1$;
\ENDWHILE
\end{algorithmic}
\end{algorithm}

\section{Average Surface Distance (ASD) Comparison}

The detailed average surface distance results of both prostate segmentation and abdominal segmentation tasks can be found in Tables~\ref{tab:multisite_abdominal_ASD} and~\ref{tab:prostate_segmentation}. the proposed CateNorm achieves the lowest average ASD on both tasks, even under the more challenging multi-domain setting.

\begin{table*}[h]
\begin{center}
\footnotesize
\setlength\tabcolsep{1pt}
\caption{ASD comparison on the abdominal datasets under the multi-domain setting (in mm). Compared with the baseline and other competitive methods, the proposed CateNorm achieves the lowest average ASD.}
\label{tab:multisite_abdominal_ASD}
\begin{tabular}{c|c|ccc|cccccc}
Method         & Forward  & BTCV          & TCIA & AVG       & Spleen        & Kid.(L)    & Gall.   & Liver         & Stom. & Panc.      \\ \shline
Baseline & BN & 1.28 & 1.17          & 1.22 & 0.59 & 0.59 & 2.36 & 0.77 & 1.93          & 1.10 \\
DSBN \cite{chang2019domain}      & BN & 1.86 & \textbf{0.90} & 1.38 & 0.51 & 0.79 & 3.07 & 0.76 & 1.96          & 1.19 \\
MS-Net \cite{liu2020ms}    & BN & 1.61 & 1.02          & 1.31 & 0.52 & 0.75 & 2.91 & 0.91 & \textbf{1.58} & 1.21 \\ \hline
\textbf{\textbf{Ours (block1)}}   & CN & \textbf{1.22} & 1.10 & \textbf{1.16} & 0.54          & 0.58          & \textbf{2.22} & \textbf{0.74} & 1.76    & 1.10          \\
\textbf{Ours (block1-4)} & CN & 1.64          & 0.97 & 1.30          & \textbf{0.51} & \textbf{0.55} & 3.25          & 0.75          & 1.75    & \textbf{1.01} \\ 
\end{tabular}
\end{center}

\end{table*}

\begin{table}[h]
\begin{center}
\setlength\tabcolsep{3pt}
\footnotesize
\caption{ASD comparison on prostate segmentation datasets under the multi-domain setting (in mm). Compared with the baseline and other competitive methods, the proposed CateNorm achieves the lowest average ASD.}
\label{tab:prostate_segmentation}
\begin{tabular}{c|c|ccc|c}
Method       & Norm        & ISBN-R & ISBN-B & I2CVB & AVG  \\ \shline
Baseline     & BN        & 0.64   & 0.71   & 1.22  & 0.86 \\
DSBN \cite{chang2019domain}         & BN        & 0.56   & 0.69   & 1.17  & 0.81 \\
MS-Net \cite{liu2020ms}        & BN        & 0.58   & 0.70   & 1.32  & 0.87 \\ 
\hline
\textbf{Ours (block1)} & CN  & 0.63   & 0.66   & 1.27  & 0.85 \\
\textbf{Ours (block1-4)} & CN  & \textbf{0.54} & \textbf{0.64} & \textbf{1.13} & \textbf{0.77} \\ 
\end{tabular}
\end{center}
\end{table}

\label{sec:asd}

\begin{figure}[t]    
\setcounter{figure}{2}
\begin{center}
  \includegraphics[width=0.8\linewidth]{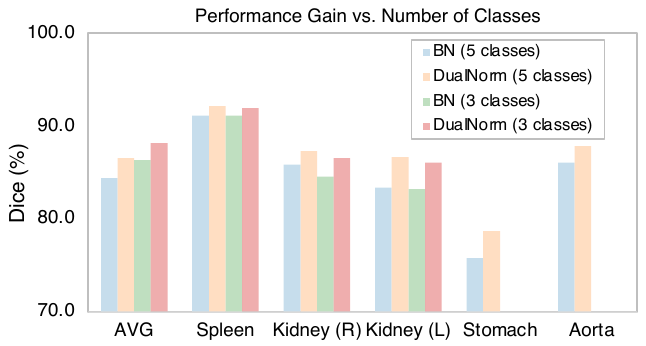}
\end{center}
  \caption{\textbf{Performance gain under partial annotation.} We compare our method to the baseline with fewer annotated classes (\ie 3/5).
  We can see that by partitioning the images into different number of regions, CateNorm consistently achieves better results than BN for all tested organs. This suggests that our algorithm is not sensitive to the number of regions.}
\label{fig:less_classes}
\end{figure}

\begin{figure}[t]

\begin{center}
  \includegraphics[width=0.7\linewidth]{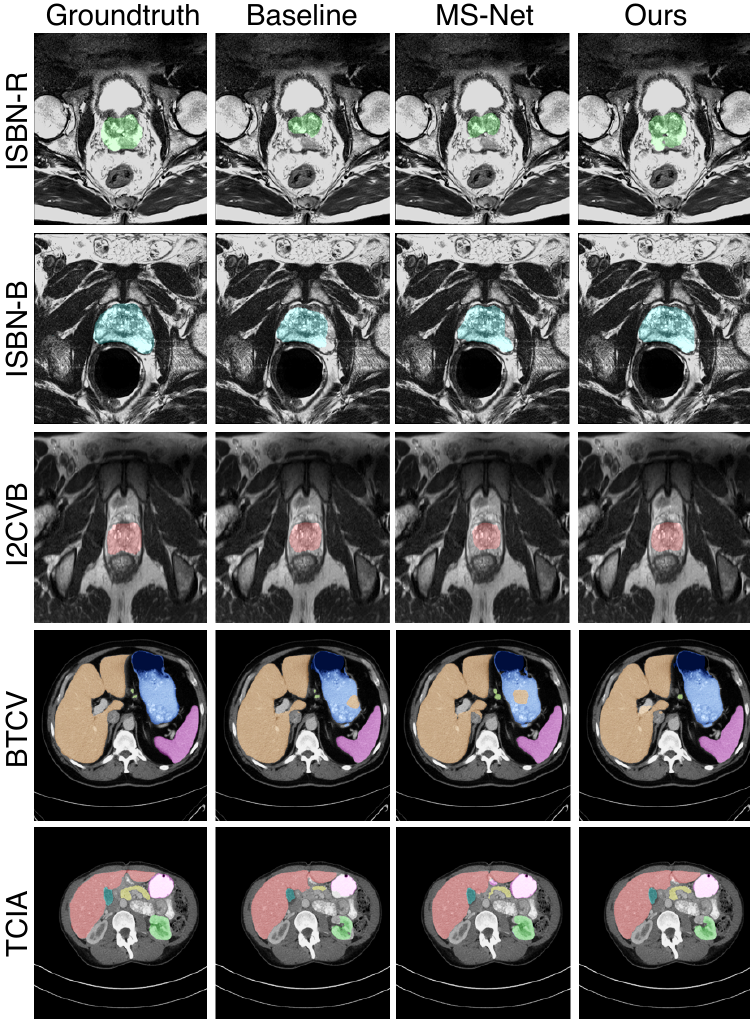}
\end{center}
  \caption{\textbf{Qualitative results comparison.} We compare our baseline and the other SOTA method under the multi-domain  setting on prostate segmentation and abdominal multi-organ segmentation. Results in the first three rows clearly show that our method outperforms others as their results are cracked and incomplete with these unapparent prostate boundaries. And the results in the last two rows show our methods could better suppress inconsistent class information inside a close segmented area (\eg reducing false positives inside the stomach) and  predict hard organs like the pancreas more accurately by incorporating general and categorical statistics.}
\label{fig:Qualitative}
\end{figure}

\begin{sidewaystable}[t]
\setcounter{table}{3}
\begin{center}
\tablestyle{1pt}{1.01}
\footnotesize
\caption{Comparison on the multi-organ segmentation dataset (BTCV) with single-domain setting (Dice Score in \%).}
\label{tab:supp_compare_abd}
\begin{tabular}{c|ccc|c|cccccccc}
Method & Pretrained & Aug & Norm & AVG & Spleen & Kidney (R) & Kidney (L) & Gallbladder & Pancreas & Liver & Stomach & Aorta \\ \shline
Baseline & \xmark & \xmark & BN       & 75.56 & 92.00 & 81.92 & 84.45 & 50.76 & 48.89 & 95.07 & 68.25 & 83.16 \\
Baseline & \xmark & \cmark & BN       & 77.83 & 91.86 & 84.36 & 86.99 & 53.92 & 51.18 & 95.11 & 75.24 & 83.97 \\ \hline
Baseline & \cmark & \cmark & BN       & 78.98 & 93.43 & 84.68 & 87.98 & 54.54 & 55.14 & 95.35 & 76.19 & 84.49 \\
Baseline & \cmark & \cmark & IN       & 78.96 & 91.93 & 83.70 & 87.83 & 54.29 & 55.08 & 95.23 & 78.72 & 84.93 \\
Baseline & \cmark & \cmark & GN       & 78.50 & 89.84 & 85.64 & 86.26 & 55.14 & 55.70 & 94.71 & 75.98 & 84.71 \\ \hline
\textbf{Ours(blk1)}    & \cmark & \cmark & CN & 79.33 & 93.07 & 85.66 & 88.26 & 52.79 & 54.34 & \textbf{95.64} & 77.38   & \textbf{87.49} \\
\textbf{Ours(blk-4)}     &
  \cmark &
  \cmark &
  CN &
  \textbf{80.37} &
  \textbf{94.63} &
  \textbf{86.29} &
  \textbf{88.64} &
  \textbf{55.51} &
  \textbf{55.91} &
  \textbf{95.64} &
  \textbf{79.80} &
  86.52 \\ 
\end{tabular}
\end{center}
\end{sidewaystable}

\begin{sidewaystable}[ht]
\begin{center}
\footnotesize
\caption{Organ-wise results on the multi-organ segmentation datasets under the multi-domain setting (Dice Score in \%). }
\label{tab:supp_multisite_abdominal}
\begin{tabular}{c|c|ccc|cccccc}
Method       & Norm  & BTCV  & TCIA  & AVG & Spleen & Kidney (L)       & Gallbladder & Liver          & Stomach & Pancreas \\ \shline
Baseline     & BN       & 82.64   & 87.33   & 84.98   & 95.22  & 93.54          & 69.63       & 96.01          & 85.52   & 69.97    \\
DSBN        & BN       & 82.67   & 87.83   & 85.25   & 95.42  & 93.49          & 69.98       & \textbf{96.16} & 86.11   & 70.34    \\
MS-Net     & BN       & 82.17   & 87.85   & 85.01   & 95.36  & 93.10          & 68.38       & 95.75          & 86.77   & 70.69    \\ \hline
\textbf{Ours (block1)} & CateNorm & 83.28   & 88.22   & 85.75   & 95.45  & \textbf{93.63} & 70.79       & 96.13          & 87.18   & 71.32    \\
\textbf{Ours (block1-4)} & CateNorm & \textbf{83.45} & \textbf{88.38} & \textbf{85.92} & \textbf{95.55} & 93.48 & \textbf{71.53} & 96.13 & \textbf{87.20} & \textbf{71.60} \\ 
\end{tabular}
\end{center}
\end{sidewaystable}

\begin{table}[htbp]
\footnotesize
\setlength\tabcolsep{3pt}
\centering
\caption{ \textbf{CateNorm is compatible to other segmentation models.}  This table compares performance on multi-domain multi-organ and prostate segmentation with DeepLabv3+~\cite{chen2018encoder} architecture. Our CateNorm consistently outperforms BN.}
\label{tab:deeplab}
\begin{tabular}{cccccc}
Backbone  & Norm     & AVG     & BTCV    & TCIA                           &                               \\ 
\shline
DeepLabV3+ & BN       & 84.58 & 81.33 & 87.83                        &                             \\
DeepLabV3+ & CN & \textbf{85.42} & \textbf{82.13} & \textbf{88.72}                        &                              \\\hline
          &          & AVG     & ISBN-R  & ISBN-B & I2CVB \\
 \shline
DeepLabV3+ & BN       & 88.54 & 90.73 & 89.35                        & 85.55                      \\ 
DeepLabV3+ & CN & \textbf{89.21} & \textbf{91.26} & \textbf{89.78}                        & \textbf{86.59} \\
\end{tabular}
\end{table}

\begin{table}[t]
\footnotesize
\begin{center}
\caption{\textbf{CateNorm is not sensitive to the warmup length.}. This table reports average accuracy (\%) of our CateNorm under deteriorated pretrained models with fewer pretraining iterations. We reduce the warmup iterations to 450, 1440, and 9000 for multi-domain prostate segmentation experiments, to investigate how our CateNorm performs when warmuped with less iterations.}
\label{tab:pretrain}
\begin{tabular}{cccc}
Warmup Iters & 450     & 1440    & 9000    \\ 
\shline
Warmup Acc.  & 73.91\% & 82.32\% & 90.10\% \\
Ours Acc.      & \textbf{91.16\%} & \textbf{91.17\%} & \textbf{91.48\%}  \\
\end{tabular}
\end{center}
\end{table}

\begin{figure}[t]
\begin{center}

  \includegraphics[width=0.9\linewidth]{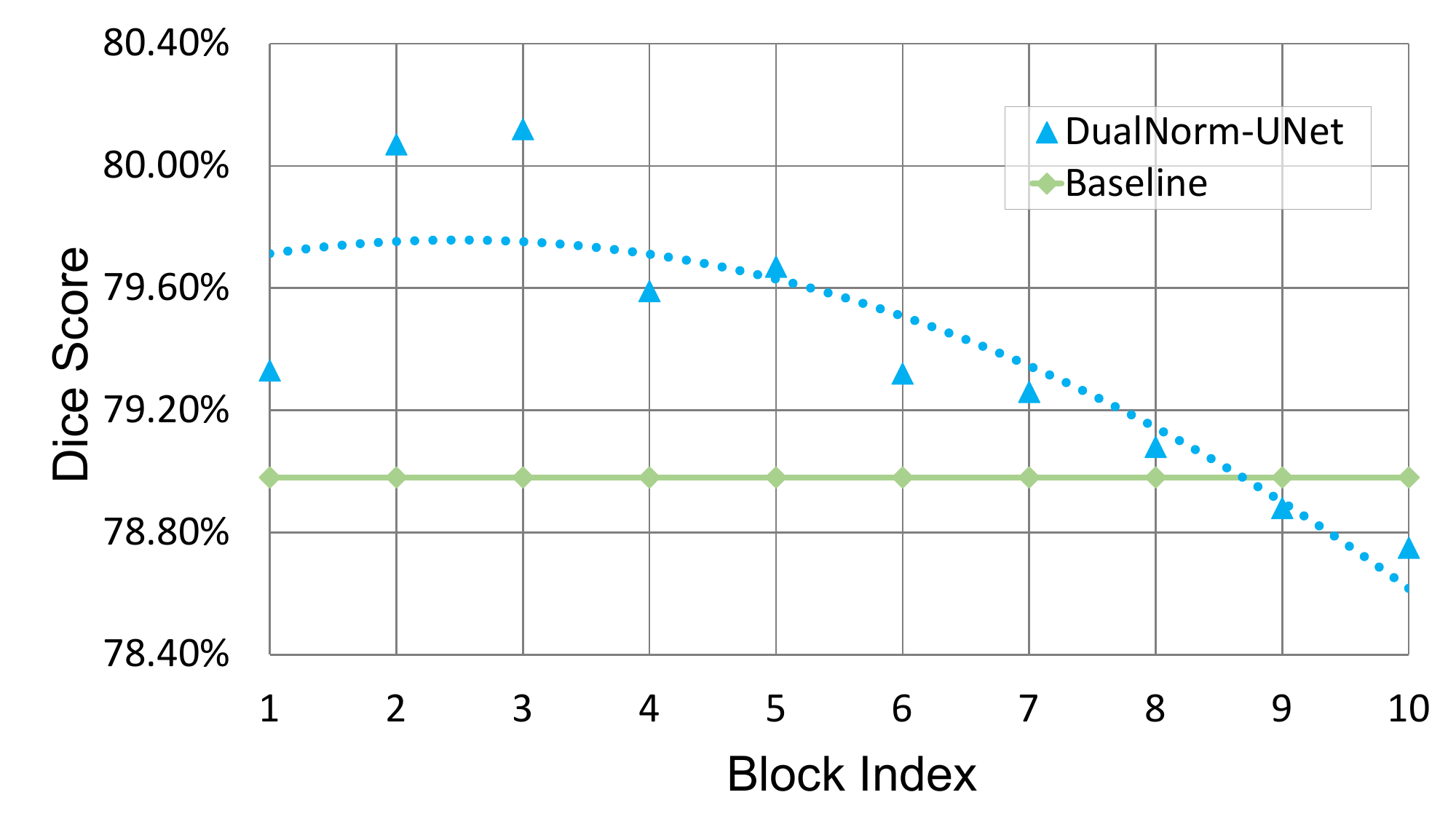}
\end{center}
  \caption{\textbf{Set CateNorm block(s) early.} This table compares performance with single CateNorm block set in different positions. Adding the CateNorm to the encoder (block index 1-5) always yields better performance than adding to the decoder (block index 6-10). In general, the performance decreases as the block index increases. We believe that it is because the earlier layers in the encoder extract lower-level features that are less discriminative than the decoder features. }
\label{fig:where_abdominal}

\end{figure}

\begin{figure}[t]    
\begin{center}

  \includegraphics[width=0.85\linewidth]{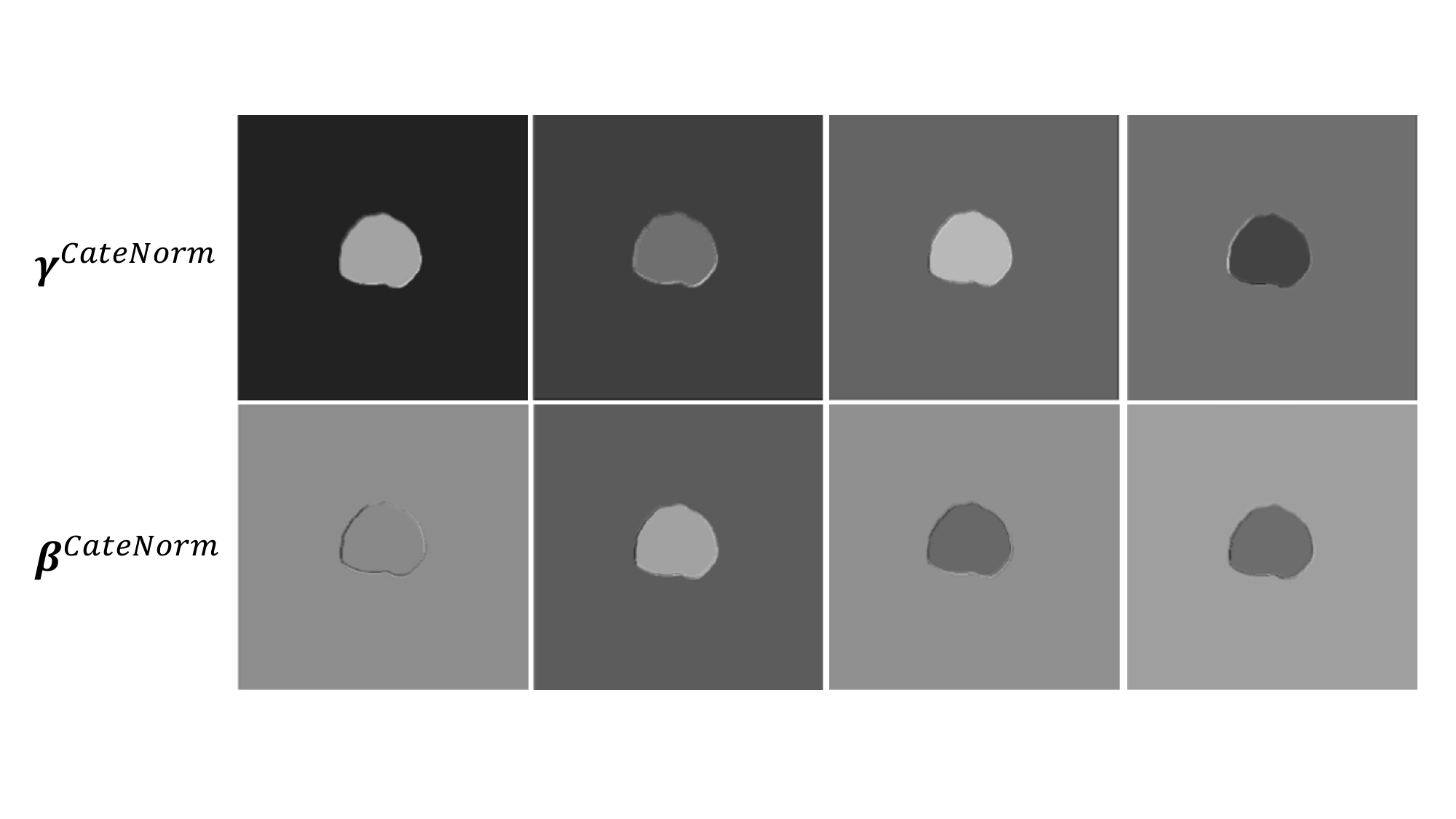}
\end{center}
  \caption{\textbf{CateNorm does normalize with semantic information.}  This figure visualizes the learned $\gamma^{\textup{CateNorm}}$ (1st row) and $\beta^{\textup{CateNorm}}$ (2nd row) of a CateNorm layer in a CateNorm block on different channels of the intermediate CateNorm layer during the second forward.  With prior class information as guidance, CateNorm can modulate spatially-adaptive parameters. Such spatial-wise modulation can be complementary to the channel-wise modulation accomplished by BN, and derives more discriminative features that benefit segmentation.}
\label{fig:Visualize_norm}
\end{figure}

\end{document}


\title{CateNorm: Categorical Normalization for Robust Medical Image Segmentation}
\titlerunning{Categorical Normalization}
%
\author{Anonymous Authors}
%
\institute{}
\maketitle              

\appendix

\section{Details of Aligning Input Distribution Algorithm}
\label{sec:align}

Assume that we have $N$ source domains $S_1, S_2, S_3, ..., S_N$, with $M_1, M_2, M_3, ..., M_N$ examples respectively, where the $i$-th domain source domain $S_i$ consists of an image set $\{\mathbf{x}_{i,j}\in \mathbb{R}^{D_{i,j}}\}_{j=1,...,M_i}$ as well as their associated annotations.
Our goal is to align the image distributions of these source domains with the target domain $T$ based on the class-wise (region-wise) statistics. 
The algorithm can be illustrated as the following steps:
\vspace{2mm}

\noindent
\textbf{Step 1: Calculate class-wise statistics of each case}

Firstly, we calculate the mean and standard deviation of each case in both the source domain and the target domain.
\begin{align}
\centering
\mu_{i,j}^{c} &= \frac{\sum_{k=1}^{|D^c_{i,j}|}\mathbf{x}^c_{i,j,k}}{|D^c_{i,j}|} , \\
\sigma_{i,j}^{c} &= \sqrt{\frac{1}{ |D^{c}_{i,j}|} \sum_{k=1}^{|D^{c}_{i,j}|}(\mathbf{x}^c_{i,j,k} - \mu_{i,j}^{c})^{2}},
\end{align}
where $\mathbf{x}_{i,j}^{c}$ denotes the pixels which belong to the $c$-th class (region) in image $\mathbf{x}_{i,j}$, with the number of pixels denoted as $|D^c_{i,j}|$. 
As a special case, $i=T$ indicates the target domain.
\vspace{2mm}



\noindent
\textbf{Step 2: Estimate aligned (new) class-wise statistics  }

Next, we calculate the mean of the statistics over all examples obtained in each domain as follows:
\begin{align}
\centering
\label{eq:stat_of_mean}
\Bar{\mu}_{i}^{c} &=\frac{\sum_{j=1}^{M_i}\mu_{i,j}^{c}}{M_i}, \\
\Bar{\sigma}_{i}^{c} &=\frac{\sum_{j=1}^{M_i}\sigma_{i,j}^{c}}{M_i}.
\end{align}

Based on the $\Bar{\mu}_{i}^{c}$, we now estimate the new class-wise mean $\Tilde{\mu}_{i,j}$ for each case of the source domain $S_i$ as follows:



\begin{align}
\begin{split}
\Tilde{\mu}_{i,j}^{c} &=\frac{\mu_{i,j}^{c} - \Bar{\mu}_{i}^{c}}{\sqrt{\frac{\sum_{j=1}^{M_i}(\mu_{i,j}^{c} - \Bar{\mu}_{i}^{c})^2}{M_i}}} \cdot  \sqrt{\frac{\sum_{j=1}^{M_T}(\mu_{T,j}^{c} - \Bar{\mu}_{T}^{c})^2}{M_T}} + \Bar{\mu}_{T}^{c}, \\
\end{split}
\end{align}
where $M_T$ denotes the number of cases in the target domain $T$.
Similarly, the new standard deviation $\Tilde{\sigma}_{i,j}$ can be computed by:

\begin{align}
\begin{split}
\Tilde{\sigma}_{i,j}^{c} &=\frac{\sigma_{i,j}^{c} - \Bar{\sigma}_{i}^{c}}{\sqrt{\frac{\sum_{j=1}^{M_i}(\sigma_{i,j}^{c} - \Bar{\sigma}_{i}^{c})^2}{M_i}}} \cdot  \sqrt{\frac{\sum_{j=1}^{M_T}(\sigma_{T,j}^{c} - \Bar{\sigma}_{T}^{c})^2}{M_T}} + \Bar{\sigma}_{T}^{c}. \\
\end{split}
\end{align}

\noindent
\textbf{Step 3: Align each case with the estimated statistics}

Based on the computed new mean and standard deviation $\Tilde{\mu}_{i,j}$, $\Tilde{\sigma}_{i,j}$, the aligned image $\Tilde{\mathbf{x}}_{i,j}$ can be computed as:
\begin{align}
\centering
\Tilde{\mathbf{x}}_{i,j}^{c} &=\frac{\mathbf{x}_{i,j}^{c} - \mu_{i,j}^{c}}{\sigma_{i,j}^{c}} \cdot \Tilde{\sigma}_{i,j}^{c} + \Tilde{\mu}_{i,j}^{c}.
\end{align}

\section{Implementation Details}

\begin{table}[htbp]
\tablestyle{6pt}{1.1}
\caption{\textbf{Data Preprocessing.}}
\begin{tabular}{c|ll}
Step & Prostate & Abdominal \\
\shline
1 & Center-cropping &  Window range Clipping [-125, 275] \\
2 & Out-of-mask slice cropping & Out-of-mask slice cropping \\
3 & Resizing & Resizing \\
4 & Z-score Normalization & Z-score Normalization

\end{tabular}

\label{tab:data_preprocess} \vspace{-.5em}
\end{table}

\begin{table}[htbp]
\caption{\textbf{Experimental Setting.}}
\tablestyle{6pt}{1.1}
\begin{tabular}{l|y{140}}
config & value \\
\shline
training iterations & 9000 \\
optimizer & Adam \\
initial learning rate & 1e-3 \\
optimizer momentum & $\beta_1, \beta_2{=}0.9, 0.999$ \\
batch size & 4 (single) 6 (multi) \\
{learning rate schedule}  & plateau scheduler  \\
Dice/CE balance factor $\lambda$ & 0.5 (abdominal) 1.0 (prostate) \\

\multirow{2}{*}{augmentation}  & horizontal flipping (prostate only) + random rotation \\
validation strategy & 5-fold \\
evaluation metric & Dice Score (\%) and ASD (mm)
\end{tabular}

\label{tab:training_setting} \vspace{-.5em}
\end{table}


\begin{algorithm}[t]
\caption{Training procedure of CateNorm}
\label{alg:training}
\begin{algorithmic}[1]
\REQUIRE Images and labels $\mathbf{x}$, $\mathbf{y}$; \\
~~~~~~~Network parameters $\theta = \{\theta_{s}, \theta_{\mathcal{N}_{B}}, \theta_{\mathcal{N}_{S}}\}$; \\
~~~~~~~Training iterations $\tau$;
\ENSURE Optimized parameters $\theta_{s}$, $\theta_{\mathcal{N}_{B}}$, $\theta_{\mathcal{N}_{S}}$;

    \STATE t $\leftarrow$ 0;
    \STATE Initialize $\theta_{s}$, $\theta_{\mathcal{N}_{B}}$ with the pretrained model and randomly initialize $\theta_{\mathcal{N}_{S}}$;
    \WHILE{$t<\tau$}
        \STATE Compute the class mask $\hat{\mathbf{y}}_B$;
        \STATE $\alpha \leftarrow 1$;
        \STATE Update ${\theta_s, \theta_{\mathcal{N}_{B}}}\leftarrow \mathop{\mathrm{min}_{\theta_s, \theta_{\mathcal{N}_{B}}}}
        {\mathcal{L}_{total}}$;
        \STATE Detach $\hat{\mathbf{y}}_B$ from gradient calculation;
        \STATE Compute the class mask $\hat{\mathbf{y}}_S$;
        \STATE $\alpha \leftarrow 0$;
        \STATE Update ${\theta_s, \theta_{\mathcal{N}_{S}}(\cdot)}\leftarrow \mathop{\mathrm{min}_{\theta_s, \theta_{\mathcal{N}_{S}}}}
        {\mathcal{L}_{total}}$;
        
        \STATE $t \leftarrow t + 1$;
\ENDWHILE
\end{algorithmic}
\end{algorithm}

\section{Average Surface Distance (ASD) Comparison}

The detailed average surface distance results of both prostate segmentation and abdominal segmentation tasks can be found in Table~\ref{tab:multisite_abdominal_ASD} and~\ref{tab:prostate_segmentation}. the proposed CateNorm achieves the lowest average ASD on both tasks, even under the more challenging multi-domain setting.

\begin{table*}[h]
\begin{center}
\footnotesize
\setlength\tabcolsep{1pt}
\begin{tabular}{c|c|ccc|cccccc}
\shline  
Method         & Forward  & BTCV          & TCIA & AVG       & Spleen        & Kid.(L)    & Gall.   & Liver         & Stom. & Panc.      \\ \hline
Baseline & BN & 1.28 & 1.17          & 1.22 & 0.59 & 0.59 & 2.36 & 0.77 & 1.93          & 1.10 \\
DSBN \cite{chang2019domain}      & BN & 1.86 & \textbf{0.90} & 1.38 & 0.51 & 0.79 & 3.07 & 0.76 & 1.96          & 1.19 \\
MS-Net \cite{liu2020ms}    & BN & 1.61 & 1.02          & 1.31 & 0.52 & 0.75 & 2.91 & 0.91 & \textbf{1.58} & 1.21 \\ \hline
\textbf{\textbf{Ours (block1)}}   & CateNorm & \textbf{1.22} & 1.10 & \textbf{1.16} & 0.54          & 0.58          & \textbf{2.22} & \textbf{0.74} & 1.76    & 1.10          \\
\textbf{Ours (block1-4)} & CateNorm & 1.64          & 0.97 & 1.30          & \textbf{0.51} & \textbf{0.55} & 3.25          & 0.75          & 1.75    & \textbf{1.01} \\ \shline
\end{tabular}
\end{center}
\caption{ASD comparison on the abdominal datasets under the multi-domain setting (in mm). Compared with the baseline and other competitive methods, the proposed CateNorm achieves the lowest average ASD.}
\label{tab:multisite_abdominal_ASD}
\end{table*}

\begin{table}[h]
\begin{center}
\setlength\tabcolsep{3pt}
\footnotesize
\begin{tabular}{c|c|ccc|c}
\shline
Method       & Norm        & ISBN-R & ISBN-B & I2CVB & AVG  \\ \hline
Baseline     & BN        & 0.64   & 0.71   & 1.22  & 0.86 \\
DSBN \cite{chang2019domain}         & BN        & 0.56   & 0.69   & 1.17  & 0.81 \\
MS-Net \cite{liu2020ms}        & BN        & 0.58   & 0.70   & 1.32  & 0.87 \\ \hline
\textbf{Ours (block1)} & CateNorm  & 0.63   & 0.66   & 1.27  & 0.85 \\
\textbf{Ours (block1-4)} & CateNorm  & \textbf{0.54} & \textbf{0.64} & \textbf{1.13} & \textbf{0.77} \\ \shline
\end{tabular}
\end{center}
\caption{ASD comparison on prostate segmentation datasets under the multi-domain setting (in mm). Compared with the baseline and other competitive methods, the proposed CateNorm achieves the lowest average ASD.}
\label{tab:prostate_segmentation}
\end{table}

\label{sec:asd}

\begin{figure}[t]    
\setcounter{figure}{2}
\begin{center}
  \includegraphics[width=0.8\linewidth]{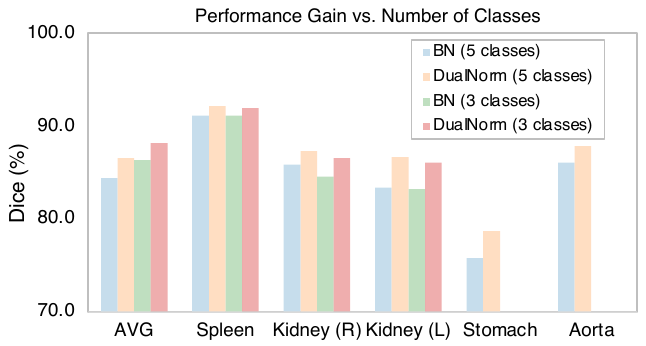}
\end{center}
  \caption{\textbf{Performance gain under partial annotation.} We compare our method to the baseline with fewer annotated classes (\ie 3/5).
  We can see that by partitioning the images into different number of regions, CateNorm consistently achieves better results than BN for all tested organs. This suggests that our algorithm is not sensitive to the number of regions.}
\label{fig:less_classes}
\end{figure}

\begin{figure}[t]

\begin{center}

  \includegraphics[width=0.7\linewidth]{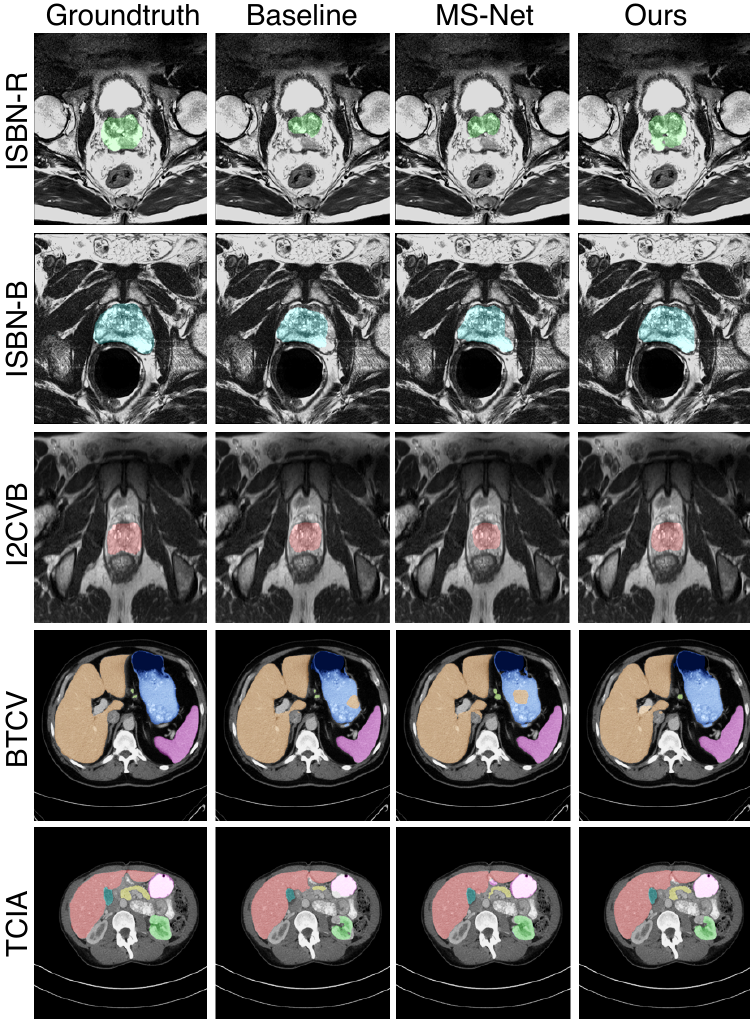}
\end{center}
  \caption{\textbf{Qualitative results comparison.} We compare our baseline and the other SOTA method under the multi-domain  setting on prostate segmentation and abdominal multi-organ segmentation. Results in the first three rows clearly show that our method outperforms others as their results are cracked and incomplete with these unapparent prostate boundaries. And the results in the last two rows show our methods could better suppress inconsistent class information inside a close segmented area (\eg reducing false positives inside the stomach) and  predict hard organs like the pancreas more accurately by incorporating general and categorical statistics.}
\label{fig:Qualitative}
\end{figure}


\begin{sidewaystable}[t]
\setcounter{table}{3}
\begin{center}
\tablestyle{1pt}{1.01}
\footnotesize

\begin{tabular}{c|ccc|c|cccccccc}
\shline  
Method & Pretrained & Aug & Norm & AVG & Spleen & Kidney (R) & Kidney (L) & Gallbladder & Pancreas & Liver & Stomach & Aorta \\ \shline
Baseline & \xmark & \xmark & BN       & 75.56 & 92.00 & 81.92 & 84.45 & 50.76 & 48.89 & 95.07 & 68.25 & 83.16 \\
Baseline & \xmark & \cmark & BN       & 77.83 & 91.86 & 84.36 & 86.99 & 53.92 & 51.18 & 95.11 & 75.24 & 83.97 \\ \hline
Baseline & \cmark & \cmark & BN       & 78.98 & 93.43 & 84.68 & 87.98 & 54.54 & 55.14 & 95.35 & 76.19 & 84.49 \\
Baseline & \cmark & \cmark & IN       & 78.96 & 91.93 & 83.70 & 87.83 & 54.29 & 55.08 & 95.23 & 78.72 & 84.93 \\
Baseline & \cmark & \cmark & GN       & 78.50 & 89.84 & 85.64 & 86.26 & 55.14 & 55.70 & 94.71 & 75.98 & 84.71 \\ \hline
\textbf{Ours(blk1)}    & \cmark & \cmark & CateNorm & 79.33 & 93.07 & 85.66 & 88.26 & 52.79 & 54.34 & \textbf{95.64} & 77.38   & \textbf{87.49} \\
\textbf{Ours(blk-4)}     &
  \cmark &
  \cmark &
  CateNorm &
  \textbf{80.37} &
  \textbf{94.63} &
  \textbf{86.29} &
  \textbf{88.64} &
  \textbf{55.51} &
  \textbf{55.91} &
  \textbf{95.64} &
  \textbf{79.80} &
  86.52 \\ \shline
\end{tabular}
\end{center}
\caption{Comparison on the multi-organ segmentation dataset (BTCV) with single-domain setting (Dice Score in \%).}
\label{tab:supp_compare_abd}
\end{sidewaystable}

\begin{sidewaystable}[ht]
\begin{center}
\footnotesize

\begin{tabular}{c|c|ccc|cccccc}
\shline  
Method       & Norm  & BTCV  & TCIA  & AVG & Spleen & Kidney (L)       & Gallbladder & Liver          & Stomach & Pancreas \\ \hline
Baseline     & BN       & 82.64   & 87.33   & 84.98   & 95.22  & 93.54          & 69.63       & 96.01          & 85.52   & 69.97    \\
DSBN        & BN       & 82.67   & 87.83   & 85.25   & 95.42  & 93.49          & 69.98       & \textbf{96.16} & 86.11   & 70.34    \\
MS-Net     & BN       & 82.17   & 87.85   & 85.01   & 95.36  & 93.10          & 68.38       & 95.75          & 86.77   & 70.69    \\ \hline
\textbf{Ours (block1)} & CateNorm & 83.28   & 88.22   & 85.75   & 95.45  & \textbf{93.63} & 70.79       & 96.13          & 87.18   & 71.32    \\
\textbf{Ours (block1-4)} & CateNorm & \textbf{83.45} & \textbf{88.38} & \textbf{85.92} & \textbf{95.55} & 93.48 & \textbf{71.53} & 96.13 & \textbf{87.20} & \textbf{71.60} \\ \shline
\end{tabular}
\end{center}
\caption{Organ-wise results on the multi-organ segmentation datasets under the multi-domain setting (Dice Score in \%). }
\label{tab:supp_multisite_abdominal}
\end{sidewaystable}

\begin{table}[htbp]
\footnotesize
\setlength\tabcolsep{3pt}
\centering
\begin{tabular}{cccccc}
\Hline
Backbone  & Norm     & AVG     & BTCV    & TCIA                           &                               \\ \hline
DeepLabV3+ & BN       & 84.58 & 81.33 & 87.83                        &                             \\
DeepLabV3+ & CateNorm & \textbf{85.42} & \textbf{82.13} & \textbf{88.72}                        &                              \\\Hline
          &          & AVG     & ISBN-R  & ISBN-B & I2CVB \\
 \hline
DeepLabV3+ & BN       & 88.54 & 90.73 & 89.35                        & 85.55                      \\ 
DeepLabV3+ & CateNorm & \textbf{89.21} & \textbf{91.26} & \textbf{89.78}                        & \textbf{86.59} \\\Hline   \end{tabular}
\caption{ \textbf{CateNorm is compatible to other segmentation models.}  This table compares performance on multi-site multi-organ and prostate segmentation with DeepLabv3+~\cite{chen2018encoder} architecture. Our CateNorm consistently outperforms BN.}
\label{tab:deeplab}
\end{table}

\begin{table}[t]
\small
\begin{center}
\begin{tabular}{cccc}\shline
Warmup Iters & 450     & 1440    & 9000    \\ \hline
Warmup Acc.  & 73.91\% & 82.32\% & 90.10\% \\
Ours Acc.      & \textbf{91.16\%} & \textbf{91.17\%} & \textbf{91.48\%}  \\ \shline

\end{tabular}
\caption{\textbf{CateNorm is not sensitive to the warmup length.}. This table reports average accuracy (\%) of our CateNorm under deteriorated pretrained models with fewer pretraining iterations. We reduce the warmup iterations to 450, 1440, and 9000 for multi-domain prostate segmentation experiments, to investigate how our CateNorm performs when warmuped with less iterations.}
\label{tab:pretrain}

\end{center}
\end{table}

\begin{figure}[t]
\begin{center}
  \includegraphics[width=0.9\linewidth]{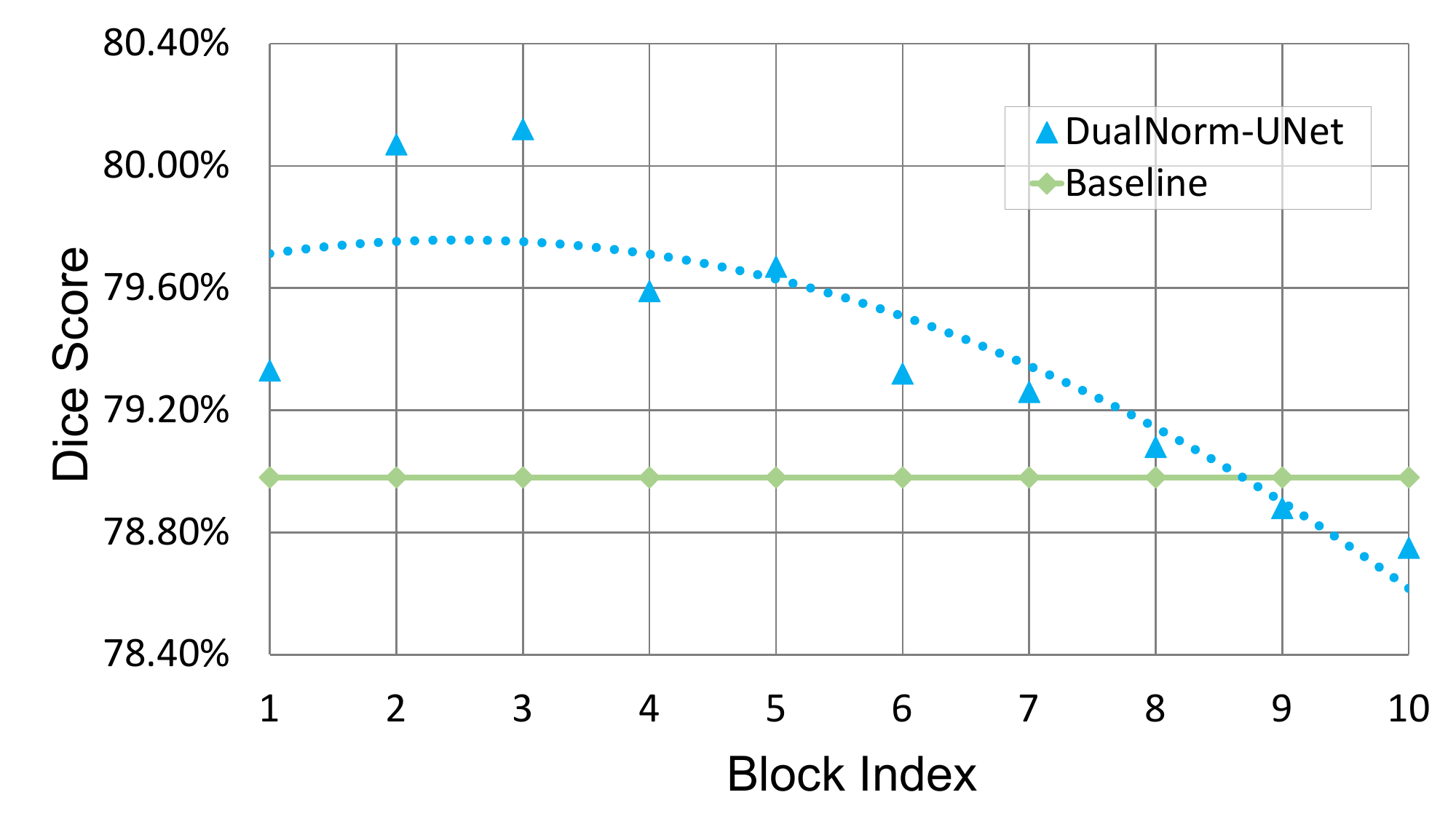}
\end{center}
  \caption{\textbf{Set CateNorm block(s) early.} This table compares performance with single CateNorm block set in different positions. Adding the CateNorm to the encoder (block index 1-5) always yields better performance than adding to the decoder (block index 6-10). In general, the performance decreases as the block index increases. We believe that it is because the earlier layers in the encoder extract lower-level features that are less discriminative than the decoder features. }
\label{fig:where_abdominal}
\end{figure}

\begin{figure}[t]    
\begin{center}
  \includegraphics[width=0.75\linewidth]{figures/gamma_beta_new2.png}
\end{center}
  \caption{\textbf{CateNorm does normalize with semantic information.}  This figure visualizes the learned $\gamma^{\textup{SPADE}}$ (1st row) and $\beta^{\textup{SPADE}}$ (2nd row) of a SPADE layer in a CateNorm block on different channels of the intermediate SPADE layer during the second forward.  With prior class information as guidance, CateNorm can modulate spatially-adaptive parameters. Such spatial-wise modulation can be complementary to the channel-wise modulation accomplished by BN, and derives more discriminative features that benefit segmentation.}
\label{fig:Visualize_norm}
\end{figure}

\clearpage
{\small
\bibliographystyle{splncs04}
\bibliography{egbib}}